\documentclass[a4paper,11pt]{article}
\usepackage{pos}
\usepackage{subfig}
\usepackage{comment}
\title{Scalar fields on fluctuating hyperbolic geometries}

\author*[a]{Muhammad Asaduzzaman}
\author[a]{Simon Catterall}

\affiliation[a]{Syracuse University,\\
Syracuse, NY 13244, USA.}

\emailAdd{masaduzz@syr.edu}
\emailAdd{smcatter@syr.edu}

\abstract{We present results on the behavior of the boundary-boundary correlation function of scalar fields propagating on discrete two-dimensional random triangulations representing manifolds with the topology of a disk. We use a gravitational action that includes a curvature squared operator, which favors a regular tessellation of hyperbolic space for large values of its coupling. We probe the resultant geometry by analyzing the propagator of a massive scalar field and show that the conformal behavior seen in the uniform hyperbolic space survives as the coupling approaches zero.
The analysis of the boundary correlator suggests that holographic predictions survive, at least, weak quantum gravity corrections. We then
show how such an $R^2$ operator might be induced as a result of integrating out massive lattice fermions and show
preliminary result for boundary correlation functions that include the effects of this fermionic backreaction on the geometry.
}

\FullConference{%
 The 38th International Symposium on Lattice Field Theory, LATTICE2021
  26th-30th July, 2021
  Zoom/Gather@Massachusetts Institute of Technology
}

\begin{document}

\maketitle

\section{Introduction} \label{sec:intro}
Regular hyperbolic lattices furnish a representation of wick-rotated anti-de Sitter spacetime. In previous
work, we have shown how the AdS-CFT correspondence can be seen by studying the propagation of free massive scalar fields on such hyperbolic lattices \cite{asaduzzamanHolographyTessellationsHyperbolic2020}. 
This earlier work confirmed
that the correlation function of the scalar fields in the boundary fall-off algebraically with boundary separation. 
This agrees with what is expected in the continuum where
the magnitude of the scaling exponent $\Delta$ of the boundary field operator is related to the bulk scalar mass, $m$, via the relation given by Klebanov \& Witten \cite{klebanovAdSCFTCorrespondence1999}
\begin{equation}
    \label{eqn_kw}
    \Delta_{\pm} = \frac{d}{2} \pm \sqrt{\frac{d^2}{4} + m^{2}}.
\end{equation}
Here, $m$ is expressed in units of the AdS curvature, while $d$ is the dimension of the boundary.  Depending on the choice of the boundary condition, either mode in the equation above can be recovered~\cite{KLEBANOV199989}. The analysis we presented earlier uses a Dirichlet boundary condition which targets the $\Delta_+$ mode.

In this work, we have taken the first step to investigate what aspects of
holography survive when fluctuations are allowed in the bulk geometry. Such fluctuations represent the
effects of quantum gravity, and since they change the continuum isometries of the bulk, they can affect the
conformal nature of the boundary theory. These effects are difficult to study in the continuum but can
be investigated in the lattice using Monte Carlo simulation. In this work, we first study the
effects of introducing a $R^2$ operator which is designed to favor regular hyperbolic lattices for strong coupling.
We then decrease the coupling to probe the structure of the boundary theory as increasingly large fluctuations are
allowed in the bulk geometry. We then argue that such an operator is naturally generated in the effective gravitational action
after integrating out massive K\"{a}hler-Dirac fermions.

The structure of the paper is described for the readers. In section \ref{sec:2d}, the methodology for studying the fluctuating hyperbolic disk is presented. Next, we present numerical results for the scaling exponents of the boundary field operators for different values of the $R^2$ coupling in \ref{sec:3d}. In the following section \ref{KD}, we formulate a model which includes dynamical fermions, and in section \ref{sec:KDresults}, we present preliminary
results for the propagator of fermions on such geometries that include the backreaction effects. Finally, we summarize our findings and discuss prospects in section \ref{sec:conclusions} .
\section{Introducing bulk curvature fluctuations} \label{sec:2d}
To simulate the effects of quantum gravity fluctuations in the bulk we
have replaced the regular tessellations studied in our
earlier work with more
general random triangulations. These are restricted so that connectivity of the 
boundary nodes is held fixed throughout and corresponds to the
connectivity of a regular $\{3,7\}$ tessellation with some fixed boundary length
while the connectivity of the inner bulk nodes is
allowed to fluctuate. This approximates the continuum notion that the metric
has negative average curvature and approaches AdS at the boundary. The amount of
fluctuation in the bulk of the disk is controlled by a discretization of the 2d gravity action supplemented with additional $R^2$ terms. 
The partition function can be written as,
\begin{equation}
    Z= \sum_{\mathcal{T}} e^{ S(k_d, \beta) }
\end{equation}
Here, the summation is carried out over all possible 2d triangulations with disk topology and fixed boundary.
The action is given by a sum over a cosmological constant
and a curvature squared term $S_Q$:
\begin{equation}
    S= k_2 N_2 +\beta S_Q
\end{equation}
The first term above is the contribution from the pure gravity action in the lattice\footnote{There is another term in the pure gravity action, but in two spacetime dimension, one of the terms is redundant.}. The modified curvature-squared term $S_Q$ takes the form
\begin{equation}
    S_Q = \sum_{i \in P_{\mathrm{bulk}}}(q_i-7)^2/q_i \label{curv_term}
\end{equation}
For large $\beta$ the coordination number of any bulk vertex
is driven to seven corresponding to the $\{3,7\}$ tessellation of the hyperbolic
disk.
\begin{figure}
    \centering
    \includegraphics[width=0.5\textwidth]{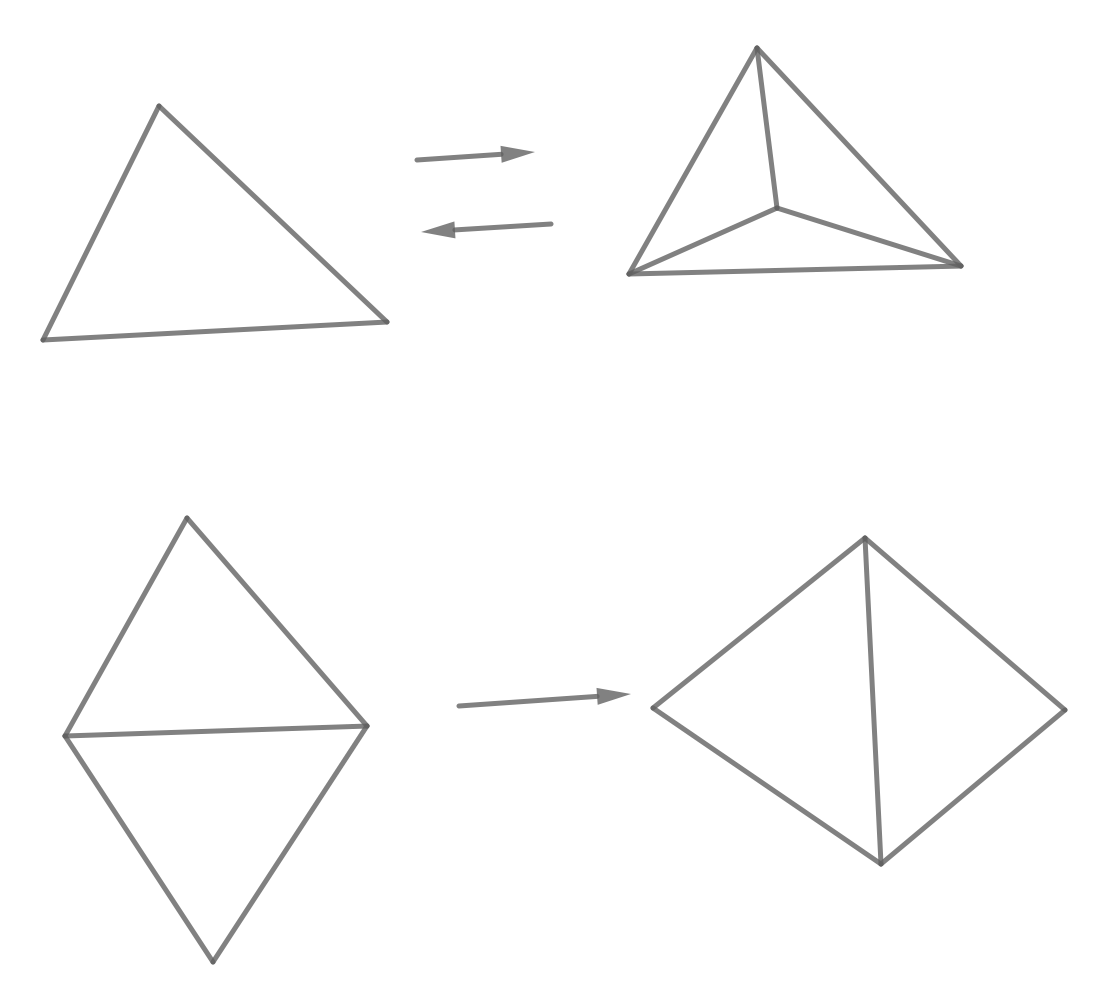}
    \caption{Ergodic moves in the class of two-dimensional triangulation. Node insertion and deletion (top), and link flip (bottom).}
    \label{fig:moves}
\end{figure}
To generate an ensemble of simplicial manifolds, a set of local moves called ``Pachner moves" are used. Any two triangulations of fixed topology are related through a sequence of Pachner moves. In two-dimensions, the Pachner moves involve inserting a node in the triangle, deleting a three fold coordinated node, and performing a link flip. The set of moves are ergodic. See Fig.~\ref{fig:moves}.
It is straightforward to use these moves to devise a simple Monte Carlo procedure 
which samples the sum over lattice geometries according to the discrete
action described above. 
In two dimensions this sum over random lattices
is known to reproduce the effects of integrating over metrics modulo diffeomorphisms
in the continuum limit. 

In practice we also introduce a volume tuning term in the simulation to allow fluctuation in the volume around a fixed target volume $V$. If in a configuration the total number of triangles in the disk is $N_2$, then the volume tuning term is
\begin{equation}
    S_{V}= \gamma (N_2-V)^2
\end{equation}

\section{Results: curvature-squared operator} \label{sec:3d}

\begin{figure}
    \centering
     \includegraphics[width=0.4\textwidth]{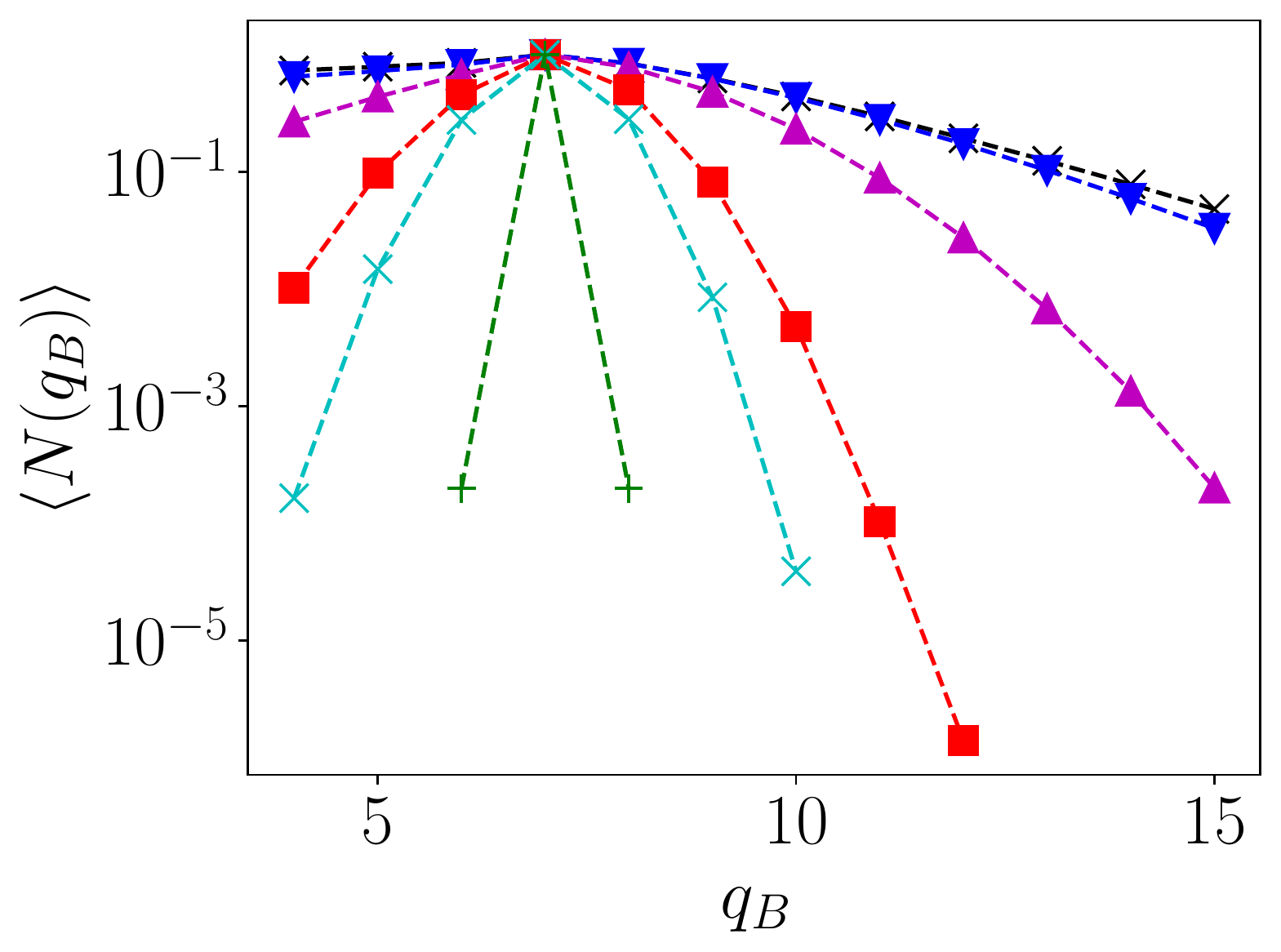}
    \caption{Number of bulk vertices $N(q_B)$ with coordination number $q$ are plotted at different $\beta$: $\beta=0.0$ (black), $\beta=0.01$ (blue), $\beta=0.1$ (purple), $\beta=0.5$ (red), $\beta=1.0$ (cyan), and $\beta=3.0$ (green).\label{fig_Nq}}
\end{figure}
\begin{figure}[!htb]
\subfloat[$\beta=0.0$\label{dp1}]{%
  \includegraphics[height=.2\textheight]{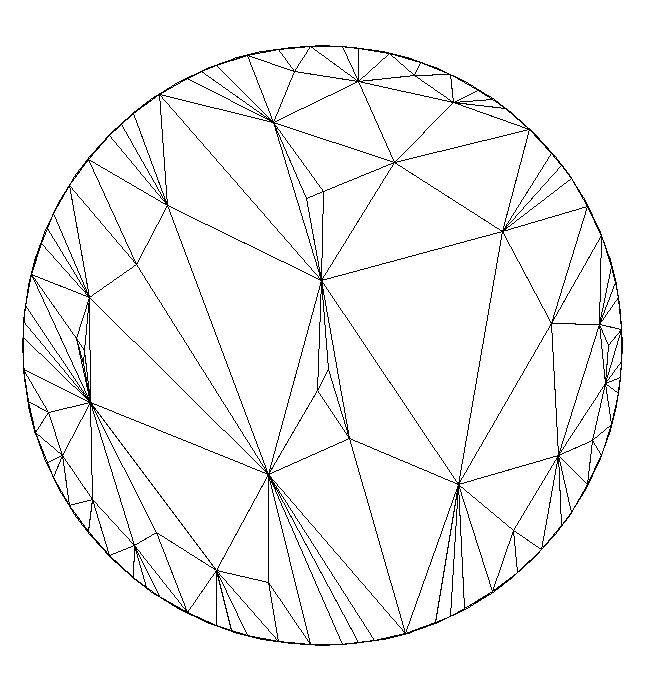}%
}\hfill
\subfloat[$\beta=0.1$\label{dp2}]{%
  \includegraphics[height=.2\textheight]{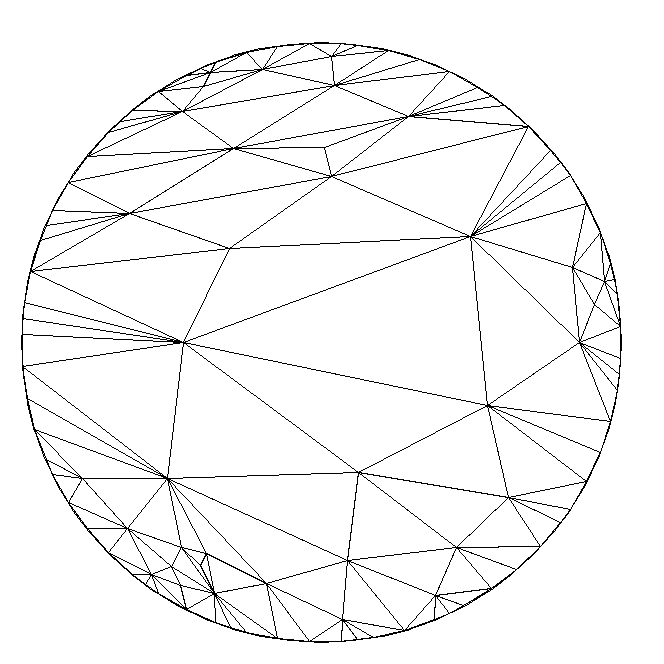}%
}
\hfill
\subfloat[$\beta=3.0$\label{dp3}]{%
  \includegraphics[height=.2\textheight]{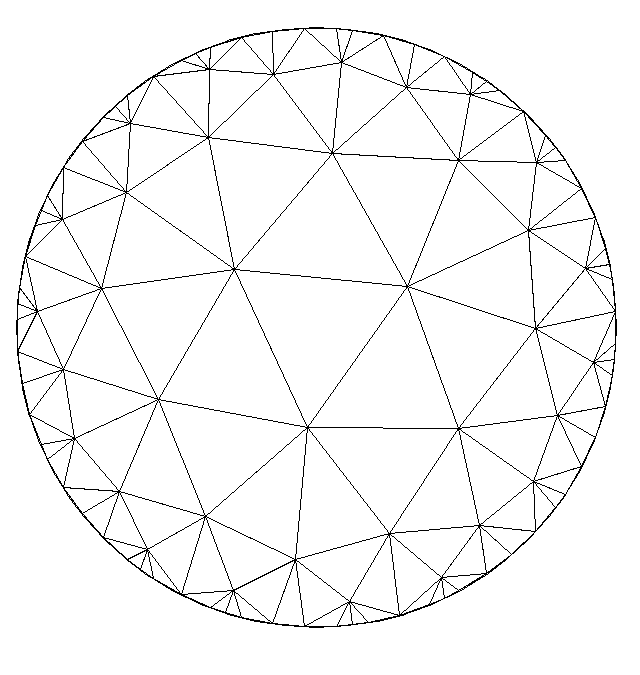}%
 }
 \caption{Sample thermalized configurations from time evolution of geometry during MCMC simulation at different $\beta$.\label{disk_snap}}
\end{figure}

\begin{figure}
    \centering
     \includegraphics[width=0.7\textwidth]{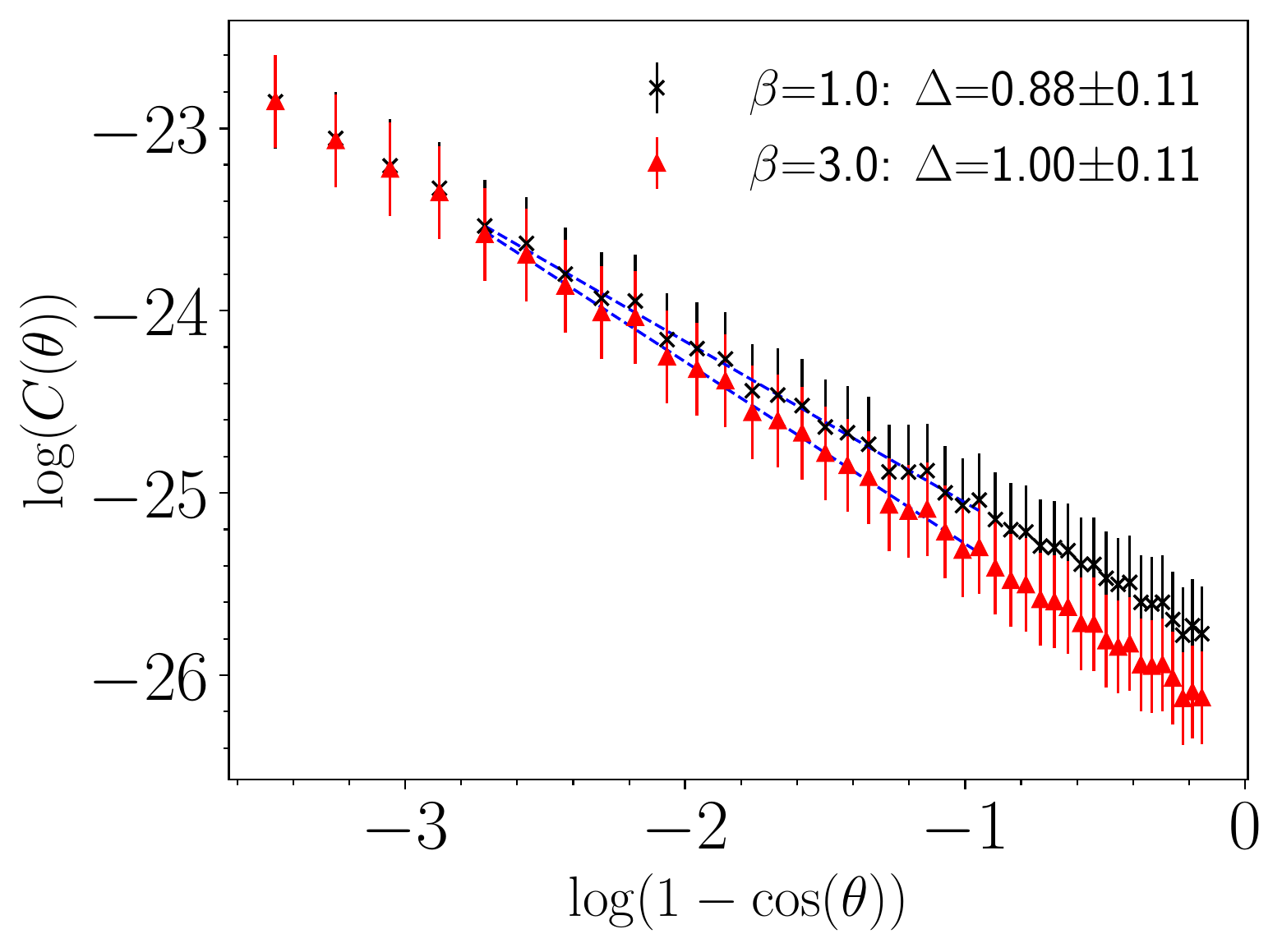}
    \caption{Two point function is plotted against boundary distance in angular coordinate for a scalar field with bulk mass $m=0$, here $r^2 \propto (1-\cos \theta)$. Slope of the fitted line denotes the scaling exponent $\Delta$ of the boundary operator. \label{corr_beta}}
\end{figure}

\begin{figure*}
\centering
\subfloat[\label{fit1}]{\includegraphics[width=0.48\textwidth]{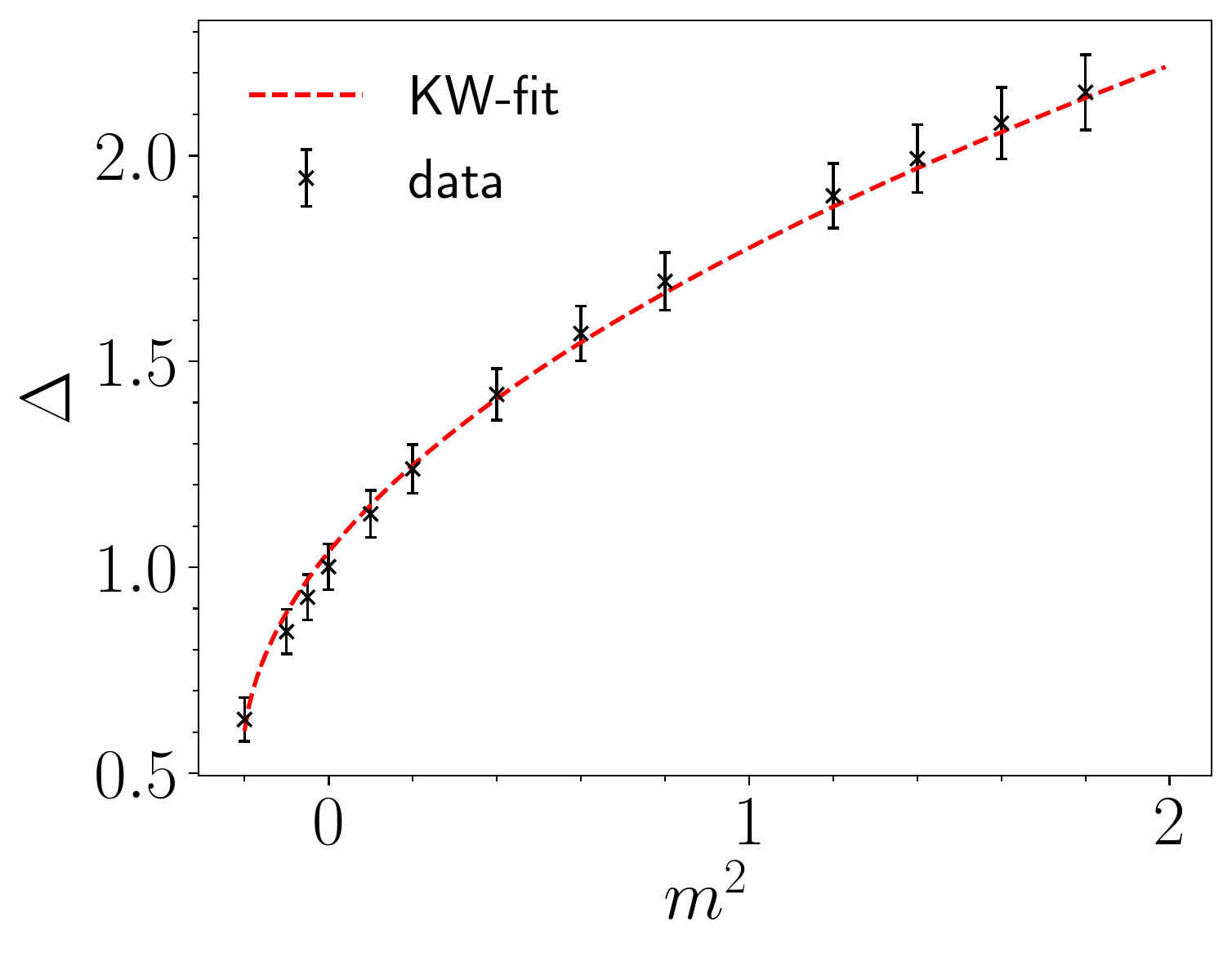}}\hfill
\subfloat[\label{fit2}]{\includegraphics[width=0.48\textwidth]{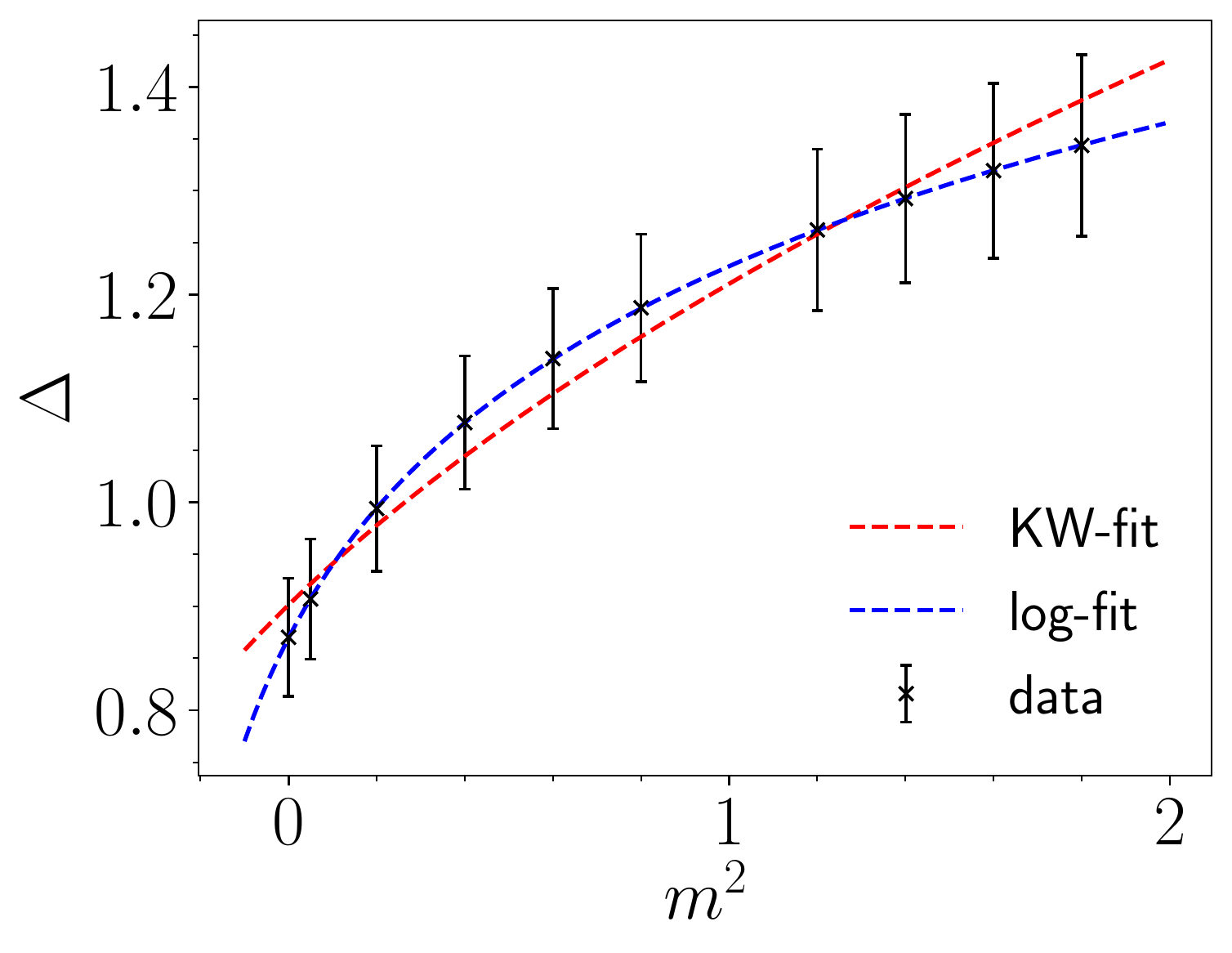}}

\caption{Bulk mass dependence of the scaling exponent at (a) $\beta=3.0$ and $\beta=1.0$.}\label{fits}
\end{figure*}

In this section, we will present our Markov Chain Monte Carlo (MCMC) simulation results for the modified $R^2$ action given in eqn.~\ref{curv_term}. We have focused
on the properties of the boundary theory and hence have examined
the boundary-boundary correlator for a massive scalar field propagating on this
geometry. To mock up the effects of a Dirichlet boundary condition on this scalar
field we have introduced a large boundary mass.
We find that the dependence of the correlators on the boundary mass vanishes as we increase the boundary mass. The analysis was repeated for several bulk masses, and similar behavior was observed for all the cases. For details of the analysis of choosing an appropriate boundary mass, see the appendix in ~\cite{asaduzzamanHolographyTessellationsHyperbolic2020}. \\

Once the boundary mass is chosen for the analysis, we investigated several collective geometry properties of the ensemble of dynamically triangulated disks. We performed a strong coupling expansion and compared the numerical results with a hand-count, giving confidence to our analysis. Fig.~\ref{fig_Nq} shows the relative number of vertices ($N_{qB}$) with a bulk coordination number $q_B$ at different coupling strength $\beta$ for target lattice volume $V=742$. The peak of the number of vertices $N_q$ is seen at seven at all nonzero coupling we considered. A fixed boundary configuration was deployed in our analysis that resembles the boundary of the $\{3,7\}$ Poincar\'e disk. Choice of this initial configuration is sufficient to ensure that the peak of the bulk coordination number is at seven. Not surprisingly, the peak is sharpened in the strong coupling limit, and we recover the geometry of the fixed tessellated disk with a high probability in our MCMC simulation. The weak coupling limit, where the peak is broad, represents stronger curvature fluctuation in the background hyperbolic manifold. 
The non-uniform curvature in the geometry can be seen from the snapshots of the geometry during MCMC simulation. Sample configurations at different couplings are shown in Fig.~\ref{disk_snap}. \\

Next, we discuss the main results of our analysis -- boundary-boundary correlator function. We compute the lattice propagators at different couplings $\beta$ and different bulk masses $m$. With geometry fluctuation in play, we find the conformal behavior of the correlators for a wide range of couplings. Fig~\ref{corr_beta} shows the boundary correlators at two different couplings for scalar mass $m=0$. Note that even at a small value of $\beta=1.0$, where $~63\%$ bulk vertices on average are 7-fold connected, we find a signature of conformal behavior. At a stronger coupling of $\beta=3.0$, where $99.6\%$ vertices are 7-fold connected, we find that the scaling exponent extracted from the linear fit of the correlator matches the Klebanov-Witten relation at Eqn.~\ref{eqn_kw}. The predicted continuum result of the boundary exponent for the zero bulk mass is $\Delta=1$ for geometry with constant negative curvature, which matches closely with our numerical result at $\beta=3.0$. 

Next, the scaling exponents for different bulk mass at a fixed $\beta$ are plotted. Results are shown for $\beta=3.0$ in the Fig.~\ref{fit1} and for $\beta=1.0$ in the Fig.~\ref{fit2}. We use two fit functions, one corresponding to
the Klebanov-Witten (KW) formula
and corresponding to Eqn.~\ref{eqn_kw}, written with three fit parameters $A$, $B$ and $L$
\begin{equation}
    \Delta=A+\sqrt{B+m^2L^2}.
\end{equation}
The other fit function we use is a logarithmic function with three fit parameters A,B,C expressed as
\begin{equation}
    \Delta=A \log (m^2+B)+C.
\end{equation}
We find that the KW relation fits well in the limit of small fluctuations, whereas the logarithmic fit describes the data at stronger fluctuation.\\

\section{Curvature fluctuation from matter fields}\label{KD}
In the previous section we presented results on the holographic behavior of a model in which the behavior
of a boundary theory is influenced by bulk quantum gravity corrections. The magnitide of the quantum corrections were
controlled by an $R^2$ operator which was added to the action by hand. It is interesting to constrict models
where such operators could be induced via coupling the system to additional dynamical matter fields. We construct
just such a model in this section.

We will see that to pick out uniform hyperbolic space we need to employ fermionic matter. To avoid the complication of having
to carry around additional structure like a spin connection and frame we have used K\"{a}hler-Dirac fermions.
The K\"{a}hler-Dirac operator squares to the (curved) space Laplacian and allows us to express the effective action
in terms of powers of the determinant of the curved space Laplacian.
Consider $N$ degenerate K\"{a}hler-Dirac (KD) fermions  with mass $m$. The partition function is given by
\begin{align}
Z&=\sum_{\mathcal{T}}\int D\phi\, e^{-k_2 N_2} \prod_{1}^{N}D\phi D\bar{\phi} e^{-\bar{\phi}^a\left(K -m\right)\phi^a} \nonumber \\
&=\sum_{\mathcal{\mathcal{T}} }{\rm det}^{\frac{N}{2}}(\widetilde{\Box})e^{-k_2 N_2}.
\end{align}
where, $\widetilde{\Box}(\mathcal{T})=-\Box(\mathcal{T})+m^2$. The effective gravitational
action can then be written as
\begin{equation}
S= \frac{N}{2} \ln \big( \mathrm{det} \; \widetilde{\Box} \big) -k_2 N_2 \label{eqn_KDaction}  
\end{equation}
It can be seen from the above expression that the Boltzmann factor is given by a non-local term. In order to find a local form of the action, we need to make some approximations. In order to do that, the Laplacian operator is factorized into three matrices
\begin{equation*}
    \widetilde{\Box} =DAD
\end{equation*}
where $D$ is a diagonal matrix with diagonal entries $d_i=\sqrt{q_i}$; and A is a matrix that depends on the connectivity of the vertices in a particular triangulation with diagonal and off-diagonal components described by 
\begin{align*}
A_{ii}&=1+m^2,\\A_{ij}&=-\frac{1}{\sqrt{q_iq_j}}C_{ij}.    
\end{align*}
In the large mass limit $m\to\infty$, A only contributes a constant factor. Hence,
\begin{equation}
\mathrm{det}\, \widetilde{\Box} \sim (\mathrm{det} D)^2 = \prod_i q_i.    \label{eqn_det}
\end{equation}
With this approximation, we arrive at a local form for the action in the large mass limit
\begin{equation} 
    S_{\rm KD}=-\frac{N}{2}\sum_{i\in\mathrm{bulk}} \ln(q_i).
\end{equation}
Thus the complete action with the K\"ahler Dirac fermion is
\begin{equation}
    S= S_{\rm KD}-k_2 N_2 + \lambda( \langle q \rangle -7 ).
\end{equation}
The last term in the equation is introduced as a constraint -- the average bulk coordination number remains seven during the evolution of the geometry in the MCMC simulation. We find from the numerical analysis that the fixed boundary geometry with identical connectivity to the $\{3,7\}$ Poincar\'{e} disk is sufficient to ensure this constraint. The logarithmic term in the effective action is then  equivalent to the curvature-squared term in Eqn.~\ref{curv_term} since    
\begin{equation*}
    \ln(q_i)\sim -\frac{1}{2\pi^2}R_i^2+\ldots  \quad.
\end{equation*}
From the discussion above, it is evident that massive K\"{a}hler Dirac fermions  dynamically generates the curvature squared term. In the $N\to\infty$ limit, regular tessellation of the Poincar\'{e} disk will be recovered with certainty in the MCMC simulation. The number of fermions $N$ plays the same role as $\beta$ in the previous construction. There may exist a transition from a holographic to a non-holographic regime as $N$ is varied. In our construction, smaller $N$ represents greater curvature fluctuation.

 \section{K\"{a}hler Dirac Operator Construction and Results \label{sec:KDresults}}
 In this section, we outline the construction of lattice K\"{a}hler Dirac fermions and present a result associated with this. For the details of the construction in the continuum and in the lattice, see this work by Banks \textit{et. el.} \cite{banksGeometricFermions1982, rabinHomologyTheoryLattice1982}. The use of the operator in the context of degenerate triangulation is discussed by Catterall \textit{et. el.} \cite{catterallKahlerDiracFermionsEuclidean2018a}.
 
The K\"ahler Dirac operator ($K$) is a square root of the Laplacian operator and 
in flat spacetime it is possible to show that each K\"ahler Dirac fermion (KD) is equivalent to two Dirac fermions in two dimensions. The equation of motion of a KD field reads
$$(d-d^\dagger+m)\Omega=0.$$
where, $\Omega=(\omega_0,\omega_\mu,\ldots ,\omega_{\mu_1\cdots\mu_D})$ is a collection of forms. 
Construction of the operator $K$ is natural in curved spacetime too, but the equivalence with Dirac spinors disappears.  It is easy to see that Laplacian operator can be written as 
$$ -\Box =K^2 =(d-d^\dagger)^2 = -d d^\dagger -d^\dagger d. $$
The above expression follows from the fundamental relation in the exterior calculus $d^2=d^{\dagger 2}=0$.\\
One of the most important properties of the K\"{a}hler-Dirac equation is that it has a natural
discretization on simplicial manifolds. Coboundary  $\bar{d}$ and boundary $\bar{\delta}=\bar{d}^T$ operators are
introduced which
are the lattice equivalent of $d$ and $d^\dagger$  respectively. The lattice operators act on lattice analogs of
the continuum p-forms which are associated with p-simplices in the lattice.  For example, the action of the $\bar{\delta}$ on a ordered list of vertices $(v_1,v_2,v_3)$ representing a triangle and carrying a 2-form field
will return an oriented linear combination of the associated links $(v_1,v_2)$, $(v_2,v_3)$ and $(v_3,v_1)$ carrying 1-form fields.
The detailed construction of the lattice K\"ahler Dirac operator can be found here \cite{catterallKahlerDiracFermionsEuclidean2018a} for the case of 4D  dynamical triangulation. 
The resultant lattice Laplacian operator is a block diagonal matrix with three non-zero blocks, each associated with different types of simplices in 2D. Thus in Eqn.~\ref{eqn_det}, there are contributions from three different matrices associated with vertices, links, and triangles.

\begin{figure}
    \centering
     \includegraphics[width=0.7\textwidth]{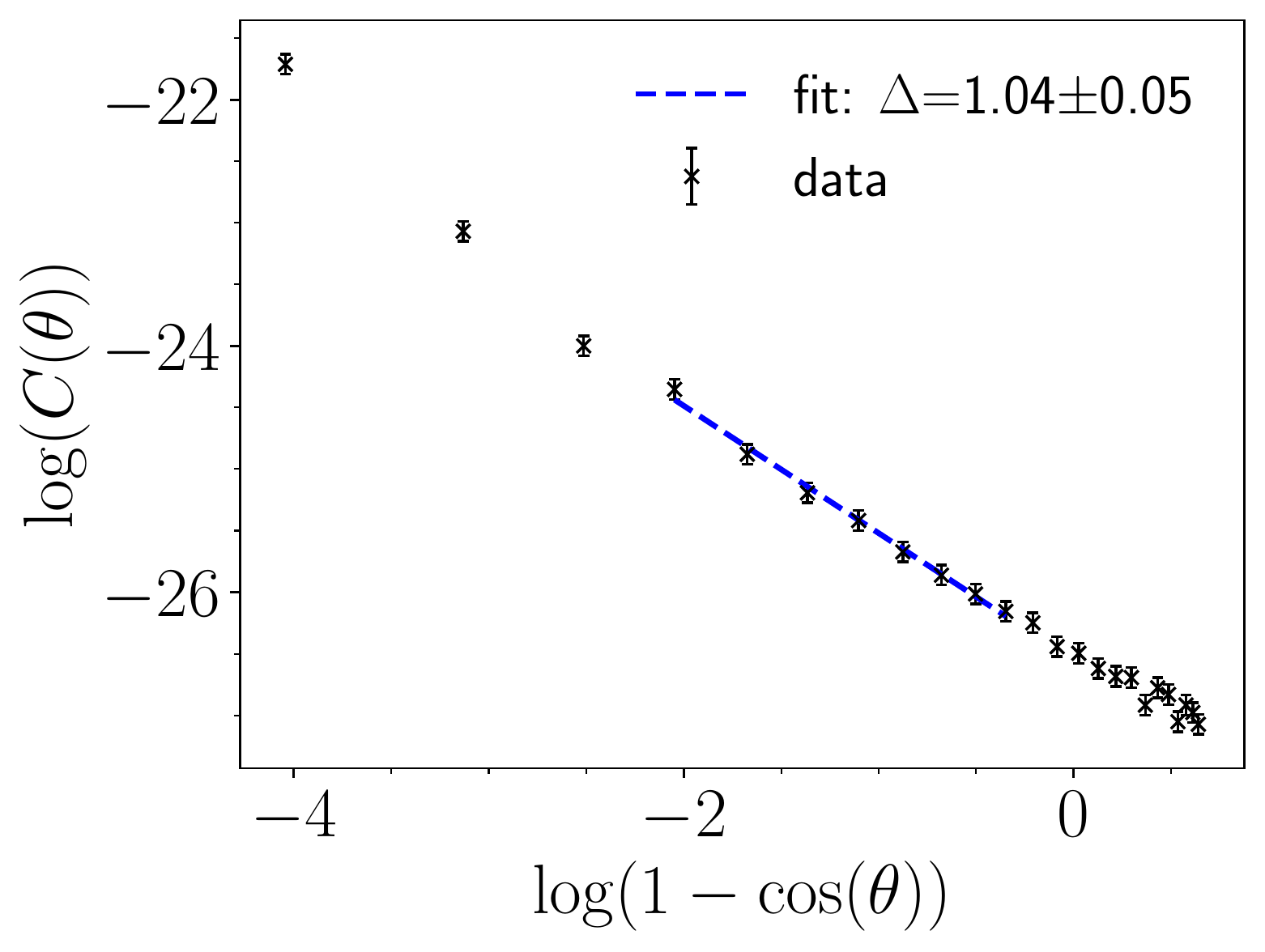}
    \caption{Lattice propagator computed from the simulation with $N_f=10$ K\"{a}hler Dirac fermions is plotted against distance-squared $r^2 \propto (1-\cos \theta)$. The slope $\Delta$ of the linear fit denotes the scaling exponent of the boundary field operator.\label{corr_KD}}
\end{figure}

We constructed the lattice K\"ahler Dirac operator to simulate a fluctuating hyperbolic disk equipped
with ten K\"ahler Dirac fermions. Resultant action is described by Eqn.~\ref{eqn_KDaction}. We found that the holographic nature of the boundary theory is reproduced in the correlator plot computed from the Markov chain Monte Carlo simulation with KD fermions. The plot is shown in Fig.~\ref{corr_KD}.
\section{Conclusions} \label{sec:conclusions}
In this lattice proceedings, we have described some preliminary results of the effects of bulk curvature fluctuations on
the holographic behavior of a two-dimensional model formulated on a space whose average curvature
is negative. We presented results from two different kinds of simulations, and they both indicate that the holographic signature survives at least weak bulk quantum gravity corrections. Further investigation is necessary to provide concrete evidence of conformality in the strongly coupled bulk quantum gravity regime. 

We also show how to couple lattice fermions to the system so that
the gravity corrections are suppressed for a large number of fields, with the
effective action picking up a curvature-squared term. It would be interesting to investigate whether a phase transition exists from the holographic to the non-holographic regime as the number of fermions $N$ is varied. Work in these directions is ongoing, and the results will be published elsewhere.
\section*{Acknowledgements} \label{sec:acknowledgements}
This work is supported in part by the U.S.\ Department of Energy (DOE), Office of Science, Office of High Energy Physics, under Award Number {DE-SC0009998}
and {DE-SC0019139}.  Numerical computations were performed at Syracuse University HTC Campus Grid: NSF award ACI-1341006.

\bibliographystyle{unsrt}

\begin{thebibliography}{1}
	
	\bibitem{asaduzzamanHolographyTessellationsHyperbolic2020}
	Muhammad Asaduzzaman, Simon Catterall, Jay Hubisz, Roice Nelson, and Judah
	Unmuth-Yockey.
	\newblock Holography on tessellations of hyperbolic space.
	\newblock {\em Physical Review D}, 102(3):034511, August 2020.
	\newblock Publisher: American Physical Society.
	
	\bibitem{klebanovAdSCFTCorrespondence1999}
	Igor~R. Klebanov and Edward Witten.
	\newblock {AdS}/{CFT} {Correspondence} and {Symmetry} {Breaking}.
	\newblock {\em Nuclear Physics B}, 556(1-2):89--114, September 1999.
	\newblock arXiv: hep-th/9905104.
	
	\bibitem{KLEBANOV199989}
	Igor~R. Klebanov and Edward Witten.
	\newblock Ads/cft correspondence and symmetry breaking.
	\newblock {\em Nuclear Physics B}, 556(1):89 -- 114, 1999.
	
	\bibitem{banksGeometricFermions1982}
	T.~Banks, Y.~Dothan, and D.~Horn.
	\newblock Geometric fermions.
	\newblock {\em Physics Letters B}, 117(6):413--417, November 1982.
	
	\bibitem{rabinHomologyTheoryLattice1982}
	Jeffrey~M. Rabin.
	\newblock Homology theory of lattice fermion doubling.
	\newblock {\em Nuclear Physics B}, 201(2):315--332, June 1982.
	
	\bibitem{catterallKahlerDiracFermionsEuclidean2018a}
	Simon Catterall, Jack Laiho, and Judah Unmuth-Yockey.
	\newblock Kähler-{Dirac} fermions on {Euclidean} dynamical triangulations.
	\newblock {\em Physical Review D}, 98(11):114503, December 2018.
	
\end{thebibliography}

\end{document}